\documentclass[a4paper,12pt]{article}

\usepackage{graphicx}
\usepackage{wrapfig}
\usepackage{grffile}
\usepackage{amssymb}
\usepackage{amsmath}
\usepackage[usenames,dvipsnames]{color}
\usepackage{verbatim}
\usepackage{hyperref}
\newcommand{\BlackBox}{\rule{1.5ex}{1.5ex}}

\makeatletter
\let\@fnsymbol\@arabic
\makeatother

\title{Cooperative Game Theoretic Solution Concepts\\for $top\text{-}k$ Problems}
\begin{document}

\author{Swapnil Dhamal\thanks{This is a work in progress. If you have any comments, suggestions, or doubts, please send an email to \texttt{swapnil.dhamal@gmailcom}}, Akanksha Meghlan\thanks{The first and second  authors have contributed equally.}, and Y. Narahari\\
\normalsize{Indian Institute of Science, Bangalore, India}
}

\date{}

\maketitle

\begin{abstract} 
The problem of finding the $k$ most critical nodes, referred to as the $top\text{-}k$ problem, is a very important one in several contexts such as information diffusion and preference aggregation in social networks, clustering of data points, etc. It has been observed in the literature that the value allotted to a node by most of the popular cooperative game theoretic solution concepts, acts as a good measure of appropriateness of that node (or a data point) to be included in the $top\text{-}k$ set, by itself.  However, in general, nodes having the highest $k$ values are not the desirable $top\text{-}k$ nodes, because the appropriateness of a node to be a part of the $top\text{-}k$ set depends on other nodes in the set. As this is not explicitly captured by cooperative game theoretic solution concepts, it is necessary to post-process the obtained values in order to output the suitable $top\text{-}k$ nodes. In this paper, we propose several such post-processing methods and give reasoning behind each of them, and also propose a standalone algorithm that combines cooperative game theoretic solution concepts with the popular greedy hill-climbing algorithm.
\end{abstract}


\section{Introduction}
\label{sec:intro}

The combinatorial problem of determining top $k$ nodes is encountered in a number of contexts, some of which are as follows:
\begin{itemize}
\item {\em Influence maximization}: determining $k$ seed nodes to trigger a campaign so as to maximize its spread in a social network \cite{kempe2003maximizing} 

\item {\em Influence limitation}: determining $k$ nodes to trigger a counter-campaign so as to limit the spread of some misinformation in a social network \cite{budak2011limiting}

\item {\em Virus inoculation}: determining $k$ nodes to vaccinate so as to restrict the spread of some epidemic in a social network \cite{abbassi2011toward}

\item {\em Preference aggregation}: determining $k$ representatives so that their aggregated preferences is close to the aggregate preference of a social network \cite{dhamal2013scalable}

\item {\em K-means clustering}: determining $k$ cluster centers for initiating the K-means clustering algorithm so as to partition a dataset into $k$ clusters in an effective way \cite{garg2013novel} (for consistency, we refer to the points in the dataset as nodes only)
\end{itemize}

In all of these problems, the input is either a network, or that can be converted to a network. 
For the problems of influence maximization, influence limitation, and virus inoculation, the input is a  weighted and directed social network, where the weight $\beta_{xy}$ (different from $\beta_{yx}$) corresponds to a measure of influence of node $x$ on node $y$.
The input to the preference aggregation problem is a weighted but undirected social network, where the weight $\beta_{xy}$ (or equivalently $\beta_{yx}$) corresponds to a measure of similarity between preferences of nodes $x$ and $y$. 
The input to the clustering problem is a dataset of points, which can be converted to a weighted undirected complete network, where the weight $\beta_{xy}$ (or equivalently $\beta_{yx}$) corresponds to a measure of similarity between data points $x$ and $y$. 

\subsection{Background and Literature}

Most of the practical $top\text{-}k$ problems are computationally hard to solve, even under very simplified models. Certain properties of the underlying objective functions, such as monotonicity and submodularity, have been exploited to develop constant-factor approximation algorithms, for example, greedy hill-climbing for the influence maximization problem under simplified models of information diffusion \cite{kempe2003maximizing}. However, these properties are lost once the models become more realistic.
So there is a need of developing algorithms that are agnostic in practice to the properties of the underlying objective function.

There exist methods for combinatorial optimization, like cross-entropy \cite{de2005tutorial}, which work well in practice. However, in recent times, cooperative game theoretic solution concepts have been used effectively in the literature for detecting top $k$ critical nodes. 
A cooperative game is represented by $(N,\nu)$, where $N$ is the set of nodes and $\nu:2^N \rightarrow \mathbb{R}$ is the valuation function.
A solution concept is an allocation $(\phi_1,\ldots,\phi_n)$, where $n=|N|$ and $\phi_x$ is the value allotted to node $x$.
There exist several solution concepts in the literature, some of the popular ones being Shapley value, Nucleolus, Banzhaf index, Gately point, etc., each possessing an individual list of certain desirable properties.
It can be seen that if the valuation function of the game is defined as the objective function itself, the value allotted to a node by most of the popular solution concepts acts as a good measure of appropriateness of that node to be included in the $top\text{-}k$ set.
This has led to application of cooperative game theory for several problems of practical interest.

Narayanam and Narahari \cite{narayanam2010shapley} propose a heuristic based on Shapley value for influence maximization in social networks, 
whereas Premm Raj and Narahari \cite{premm2012influence} use Shapley value for the problem of influence limitation in social networks.
Garg, Narahari, and Murty \cite{garg2013novel} formulate the clustering problem as a cooperative game and also show that Shapley value for the game can be exactly computed in polynomial time,
while Dhamal et al. \cite{dhamal2012pattern} prove the equivalence of several solution concepts, namely, Shapley value, Nucleolus, Gately point, and $\tau$-value for the formulated game.

\subsection{Motivation}

As mentioned earlier, it has been observed in the literature that the value allotted to a node by most of the well-known cooperative game theoretic solution concepts, acts as a good measure of appropriateness of that node to be included in the $top\text{-}k$ set. However, it can be easily seen using experiments that the value allotted to a node is as a good measure, when the node has to chosen all by itself. As the appropriateness of a node to be a part of the $top\text{-}k$ set depends on other nodes in the set, nodes having the highest $k$ values are generally not the desirable $top\text{-}k$ nodes. This fact is not explicitly captured by cooperative game theoretic solution concepts, and so in order to output the suitable $top\text{-}k$ nodes, it is necessary to post-process the obtained values. Hence the motivation behind this paper is to propose several such post-processing methods.
We also propose a standalone algorithm that combines cooperative game theoretic solution concepts with the popular greedy hill-climbing algorithm, taking into account the fact that the appropriateness of a node to be chosen in the $top\text{-}k$ set depends on other nodes in the set. 

We now present a number of post-processing methods, primarily in the context of information diffusion; they can be extended to other contexts. However, some methods would be more suitable for a given application than others.
One can derive variants of the proposed methods or use multiple methods in conjunction.

\section{Post-processing Methods}
\label{sec:methods}

Given an input that is a graph (or that can be converted to a graph), let $\beta_{xy}$ denote the weight of edge $xy$, based on the context as described before.
%
%
Starting from the null set, the $top\text{-}k$ set builds up as nodes get added to it, until it reaches cardinality $k$.
The most direct and na\"ive method of obtaining the $top\text{-}k$ set is to sort nodes in descending order of their values, say \textit{ordered\_list}, and then choose the first $k$ nodes from the list. However, as explained earlier, nodes having the highest $k$ values are generally not the desirable $top\text{-}k$ nodes, and so there is a need to post-process the obtained values in order to build an effective $top\text{-}k$ set.
We propose several post-processing methods that can be broadly classified into two types, namely,
\begin{itemize}
\item Eliminating neighbors of chosen nodes
\item Discounting values of neighbors of chosen nodes
\end{itemize}
In the following methods, neighbors of a node account for both its in-neighbors and out-neighbors; however, the accounting of neighbors can be altered based on the application.
In the methods that follow, in order to obtain an ordering over multiple nodes that are allotted equal values, the ties are broken randomly.

\subsection{Eliminating neighbors of chosen nodes}

As choosing nodes na\"ively may result in the $top\text{-}k$ nodes to be clustered in one part of the network, these methods try to choose the nodes such that they are appropriately spread in the network, so that they influence or represent as many distinct nodes as possible.
In all the methods of this type, if it is not possible to choose any more nodes using the elimination approach, the methods reiterate over \textit{ordered\_list} and na\"ively choose the unchosen nodes in order.
It is to be noted that in the methods of this type, in the context of information diffusion where the network is directed, the term {\em neighbors} refers to both in-neighbors and out-neighbors. 
Also, if a node $x$ is both in-neighbor and out-neighbor of a node $y$, we consider the mutual edge weight to be $\max \{ \beta_{xy} , \beta_{yx} \}$ whenever applicable.

\subsubsection{Eliminating neighbors of chosen nodes always}

This method is the one used in \cite{narayanam2010shapley} for influence maximization in a social network.
It keeps on choosing the nodes in order from \textit{ordered\_list} and skips a node if any of its neighbors is already chosen in the $top\text{-}k$ set. 
This method is observed to perform well in the context of information diffusion since once a node is chosen, it would more or less influence its out-neighbors either directly or indirectly (in multiple hops through other nodes); furthermore, if a node is chosen before its in-neighbors because of its high value, it is likely that its in-neighbors would have a high value only because they influence the former (a more influential node); so the method eliminates both in-neighbors and out-neighbors. 
Similarly, for the preference aggregation problem, once a node is chosen, it would more or less represent its neighbors well.
 Note that this method is unsuitable for the clustering problem as the converted input is a complete network.

\subsubsection{Eliminating neighbors of chosen nodes based on a threshold}

This method is similar to the one used in \cite{garg2013novel} in the context of clustering.
The method that eliminates neighbors of chosen nodes always, suffers from the fact that multiple nodes that are highly influential (more or less independent of each other) may be connected with low edge weights; in such cases, it is undesirable to eliminate the neighbors of such influential nodes.
So instead of eliminating neighbors of chosen nodes always, this method keeps on choosing the nodes in order from \textit{ordered\_list} and skips a node if any of its neighbors which is already chosen in the $top\text{-}k$ set, is such that the corresponding edge weight exceeds a certain threshold.
This method would work well in all the contexts, provided the threshold is chosen appropriately.
One can come up with several variants of this method; for instance, the threshold could be a fixed one for the entire network or dataset \cite{garg2013novel}, or it could be a function of the value of the chosen node itself \cite{dhamal2012pattern}.

\subsubsection{Eliminating neighbors of chosen nodes based on their local networks}

This method determines whether a node should be selected based on its local neighborhood.
It keeps on choosing the nodes in order from \textit{ordered\_list} and skips a node $x$ if there exists its neighbor $y$ which is already chosen in the $top\text{-}k$ set such that, when all the neighbors of $y$ are ordered in decreasing order of their edge weights with $y$, then $x$ lies in the {\em first  half}.
Note that this fraction {\em half} is just a natural first guess; as it acts like a threshold, it can also be a fixed one for the entire network or it can be a function of the value of the chosen node.
Intuitively, this method does not eliminate a node when it is a good candidate in the local neighborhood of the already chosen nodes, that is, it is less likely to be influenced by or to influence (or is less similar to) the nodes in the $top\text{-}k$ set.

\subsection{Discounting values of neighbors of chosen nodes}

The elimination methods are strict owing to their $0/1$ nature of eliminating a node. Moreover, it is highly likely that it would not be possible to choose any more nodes using these methods beyond a certain $k$, resulting in a na\"ive selection of the unchosen nodes. One way to overcome these problems is to discount the values of the neighbors of the chosen nodes based on certain criteria instead of eliminating them. 

These methods run in $k$ steps where the value of each node gets updated in each step $t$. Let $top\text{-}k^{(t)}$ be the $top\text{-}k$ set in step $t$, and let $\phi_x^{(t)}$ be the value of node $x$ in step $t$. The initializing $top\text{-}k$ set can be given by $top\text{-}k^{(0)} = \{\}$, and the initializing value of a node $\phi_x^{(0)} = \phi_x$ is the original value allotted to it. In the methods of this category, we do not update the values of the nodes which are already chosen in the $top\text{-}k$ set, that is,

\begin{equation}
\phi_z^{(t)} = \phi_z^{(t-1)} \;\;\; \forall 
z \in top\text{-}k^{(t-1)}
\end{equation}

It is important that the values of the chosen nodes do not change after they get added to the $top\text{-}k$ set, since the discounting of the values of unselected nodes critically depends on these values.
In order to explain the methods of this category, let $y$ be the node chosen in step $t$, that is, $top\text{-}k^{(t)} \setminus top\text{-}k^{(t-1)} = \{y\}$. Also let $\mathcal{N}_w$ be the set of neighbors of a node $w$.

\subsubsection{Discounting values of neighbors of chosen nodes - I}

\begin{equation}
\phi_x^{(t)} = (1-\beta_{yx}) \, \phi_x^{(t-1)} \;\;\; \forall x \in \mathcal{N}_y \setminus top\text{-}k^{(t-1)}
\end{equation}

\begin{equation}
\text{or} \;\;\; \phi_x^{(t)} = \left( \prod_{w \in \mathcal{N}_x \cap top\text{-}k^{(t)}} (1-\beta_{wx}) \right) \, \phi_x^{(0)} 
\end{equation}

This discounting is natural in the independent cascade model of information diffusion where $\beta_{yx}$ corresponds to the parameter $p_{yx}$ (in independent cascade model, when node $y$ first becomes active at time $\tau$, it is given a single chance to activate each of its currently inactive neighbor $x$ at time $\tau+1$ and it succeeds with probability $p_{yx}$).
Since node $x$ would get activated because of node $y$ with probability $p_{yx}$, the value of node $x$ should be discounted by the factor of $p_{yx}$ whenever any of its neighbor $y$ gets chosen in the $top\text{-}k$ set. 
Equivalently, since node $x$ would not get directly activated by any of its neighbors that are chosen in the $top\text{-}k$ set, with probability $\prod_{w \in \mathcal{N}_x \cap top\text{-}k^{(t)}} (1-p_{wx})$, the value of node $x$ is updated using this factor. 
Note that we ignore the possibility that the influence of node $y$ can reach node $x$ in multiple hops.

\subsubsection{Discounting values of neighbors of chosen nodes - II}

%

\begin{equation}
\phi_x^{(t)} = \phi_x^{(t-1)} - \beta_{yx} \, \phi_x^{(0)} \;\;\; \forall x \in \mathcal{N}_y \setminus top\text{-}k^{(t-1)}
\end{equation}

\begin{equation}
\text{or} \;\;\; \phi_x^{(t)} = \left(1 - \sum_{w \in \mathcal{N}_x \cap top\text{-}k^{(t)}} \beta_{wx} \right) \phi_x^{(0)}
\end{equation}

This discounting is natural in the linear threshold model of information diffusion where $\beta_{yx}$ corresponds to the parameter $b_{yx}$ (in linear threshold model, $b_{yx}$ is the influence weight of node $y$ on node $x$ such that the sum of the influence weights from all of its incoming neighbors is at most $1$; node $x$ gets activated if the sum of the influence weights from its active incoming neighbors exceeds a certain threshold $\theta_{x}$ that is drawn from a uniform distribution between $[0,1]$). 
Analogous to the argument in the previous method, since node $x$ would not get directly activated by any of its neighbors that are chosen in the $top\text{-}k$ set, with probability $1 - \sum_{w \in \mathcal{N}_x \cap top\text{-}k^{(t)}} b_{wx}$, the value of node $x$ is updated using this factor. 
Note that in this method also, we ignore the possibility that the influence of node $y$ can reach node $x$ in multiple hops.

\subsubsection{Discounting values of neighbors of chosen nodes - III}

\begin{equation}
\phi_x^{(t)} = \phi_x^{(t-1)} - \beta_{xy} \, \phi_y^{(t-1)} \;\;\; \forall x \in \mathcal{N}_y \setminus top\text{-}k^{(t-1)}
\end{equation}

\begin{equation}
\text{or} \;\;\; \phi_x^{(t)} = \phi_x^{(0)} - \sum_{w \in \mathcal{N}_x \cap top\text{-}k^{(t)}} \left( \beta_{xw} \, \phi_w^{(t-1)} \right)
\end{equation}

This discounting is natural in both independent cascade and linear threshold models of information diffusion (note the swapping of $x$ and $y$ with respect to the previous methods).
In the independent cascade model, as node $x$ influences node $y$ directly with probability $p_{xy}$, it gets a fractional share of the value of node $y$ (since $x$ would be influencing other nodes indirectly, through $y$).
Now given that node $y$ is chosen in the $top\text{-}k$ set, the share of $y$'s value should be removed from the value of $x$. 
This method uses a simplified expression for this share, namely, $p_{xy} \, \phi_y^{(t-1)}$.
Similar argument leads this share to be $b_{xy} \, \phi_y^{(t-1)}$ in the linear threshold model.
Note that we may be possibly removing more share than required since there may exist multiple neighbors of $x$, that are already chosen in the $top\text{-}k$ set, with shares of similar nature (for example, they may be likely to influence almost the same set of nodes).  Owing to this, it is possible for the value of a node to become negative.
\\


Furthermore, depending on the application, one may also update the values of the nodes using a suitable convex combination of the aforementioned discounting methods.

\section{Combining with greedy hill-climbing algorithm}

The greedy hill-climbing algorithm is one of the most basic methods used for combinatorial optimization. 
Starting with the null set, the greedy hill-climbing algorithm selects $k$ nodes one at a time, each time choosing a node that provides the largest marginal increase in the value of the objective function.

The following method directly utilizes the fact that the appropriateness of a node to be chosen in the $top\text{-}k$ set depends on other nodes which are already chosen in the set. It is to be noted that this method is a standalone algorithm and not a post-processing method. 

This method runs in $k$ steps where in step $t$, we define a new cooperative game $(N \setminus top\text{-}k^{(t-1)} , \omega^{(t)})$ where the valuation function $\omega^{(t)} : 2^{N \setminus top\text{-}k^{(t-1)}} \rightarrow \mathbb{R}$ is given by
\begin{equation}
\omega^{(t)}(S) = \nu(top\text{-}k^{(t-1)} \cup S) - \nu(top\text{-}k^{(t-1)})
\end{equation}
Now form $top\text{-}k^{(t)} = top\text{-}k^{(t-1)} \cup y$, where $y$ is a node with the maximum allocation value in the game $(N \setminus top\text{-}k^{(t-1)} , \omega^{(t)})$.

\subsection{Combining Shapley value with greedy hill-climbing}

The Shapley value of a node can be computed as the average marginal contribution of the node in all possible permutations over the set of nodes. Though computing exact Shapley value allocation for a general game is hard, it can be approximately computed by considering only a subset of the set of all permutations. 

We present a simple algorithm for using Shapley value with greedy hill-climbing for $top\text{-}k$ problems. The first node to be chosen in the $top\text{-}k$ set (say $y_1$) can be obtained as the node with the highest Shapley value, where the Shapley value allocation is computed by considering only a subset of the set of all permutations. 
The second node to be chosen in the $top\text{-}k$ set (say $y_2$) can be obtained as the node with the highest Shapley value among the set of nodes excluding $y_1$, where the Shapley value allocation is now computed by considering only a subset of the set of all permutations in which $y_1$ is positioned in the first place. 
In general, the $\kappa^{th}$ node to be chosen in the $top\text{-}k$ set (say $y_{\kappa}$) can be obtained as the node with the highest Shapley value among the set of nodes excluding $y_1, \ldots, y_{\kappa-1}$, where the Shapley value allocation is computed by considering only a subset of the set of all permutations in which $y_1, \ldots, y_{\kappa-1}$ are positioned in the first $\kappa-1$ places. 

Note, however, that the running time of the combined algorithm would be approximately $k$ times the running time of computing the approximate Shapley value allocation. Some of the possible ways of reducing the running time would be to:
\begin{enumerate}
\item[(a)] reduce the number of permutations for the computation of approximate Shapley value allocation, since we are only interested in obtaining the node with the highest or near-highest value, and not in the actual allocation itself,
\item[(b)] reduce the number of permutations for later steps, that is, consider more permutations for obtaining $y_1$ than that for obtaining $y_2$, and so on, owing to the inherent fact that obtaining an effective $y_1$ is more critical than obtaining an effective $y_2$,
\item[(c)] use the combined algorithm only for choosing the first few $\tilde{k}$ critical nodes and then use some post-processing or discounting approach for choosing the remaining $k-\tilde{k}$ nodes,
\item[(d)] store the marginal contributions of nodes in permutations which are likely to reoccur, for instance, permutations in which nodes with highest weighted degrees are positioned in the first few places.
\end{enumerate}

\section{Discussion}

We proposed several methods of post-processing the values obtained using any given solution concept from cooperative game theory. The proposed methods are quite generic in nature, but it is possible that some methods better suit a given context than others. One can derive variants of the proposed methods or use multiple methods in conjunction.
We also proposed a standalone algorithm that combines cooperative game theoretic solution concepts with the greedy hill-climbing algorithm, by considering that the appropriateness of a node to be chosen in the $top\text{-}k$ set depends on other nodes in the set. 

We compared the proposed methods on small datasets for the purpose of our observations, the results of which are not presented here. The results of extensive experimentation on large datasets will be directly added in the subsequent versions of the paper.

For any practical problem, the natural valuation function would lead to intractable computation for most solution concepts. Though approximate algorithms exist for several solution concepts, an alternative would to formulate a valuation function that closely resembles the problem and at the same time, facilitates efficient computation of solution concepts. 

%

\bibliographystyle{plain}
\bibliography{cgt_topk_dhamal_references}

\end{document}